\documentclass[runningheads,final]{llncs}
\usepackage[commandnameprefix=always,todonotes={textsize=small,obeyFinal}]{changes}
\usepackage[T1]{fontenc}
\usepackage{siunitx}
\usepackage[utf8]{inputenc}
\usepackage{graphicx}
\usepackage{booktabs}
\usepackage{listings}
\usepackage{xcolor}
\usepackage{url}
\usepackage{hyperref}
\usepackage{microtype}

\definecolor{codegray}{rgb}{0.95,0.95,0.95}
\definecolor{codekw}{rgb}{0.10,0.20,0.55}
\definecolor{codecomment}{rgb}{0.35,0.45,0.35}
\begin{document}

\title{Unifying von-Neumann HPC and\linebreak[3]{} Neuromorphic Acceleration via the\linebreak[3]{} EBRAINS Research Infrastructure:\linebreak[3]{} A Framework for High-Performance Workflows}

\titlerunning{Unifying von-Neumann HPC and Neuromorphic Acceleration via EBRAINS}

\author{%
  Krishna Kant Singh\inst{1}\orcidID{0000-0002-7472-6299} \and
  Charl Linssen\inst{1}\orcidID{0000-0002-8140-2866} \and
  Eric M{\"u}ller\inst{2}\orcidID{0000-0001-5880-2012} \and
  Eleni Mathioulaki\inst{3}\orcidID{0009-0008-2019-7724} \and
  Wouter Klijn\inst{1}\orcidID{0000-0002-8996-488X} \and
  Lena Oden\inst{4}\orcidID{0000-0002-9670-5296}
}

\authorrunning{Singh et al.}

\institute{%
  Simulation \& Data Lab Neuroscience, J{\"u}lich Supercomputing Centre, Forschungszentrum J{\"u}lich, Germany\linebreak[2]{}
  \email{\{k.singh, c.linssen, w.klijn\}@fz-juelich.de}
  \and
  Kirchhoff Institute for Physics, Ruprecht-Karls-Universit{\"a}t Heidelberg, Germany\linebreak[2]{}
  \email{mueller@kip.uni-heidelberg.de}
  \and
  ATHENA Research \& Innovation Center, Marousi, Greece\linebreak[2]{}
  \email{emathioulaki@athenarc.gr}
  \and
  Computer Engineering, FernUniversit\"{a}t in Hagen, Germany\linebreak[2]{}
  \email{lena.oden@fernuni-hagen.de}
}


\maketitle

\begin{abstract}
Modern scientific workflows increasingly span diverse computing architectures, yet executing a single computational model across disparate systems often forces researchers to maintain fragmented, site-specific pipelines.
In this paper, we address this challenge within the domain of computational neuroscience by presenting a unified, cloud-based workflow orchestrated via EBRAINS JupyterLab.
This workflow enables users to transparently execute spiking neural networks on both von-Neumann supercomputers and neuromorphic hardware.
Using a single federated identity, the system dispatches jobs to HPC sites (JUSUF, Galileo100) via PyUNICORE and to the SpiNNaker-1 neuromorphic system via the Neuromorphic Computing Platform Interface\chdeleted{ (NMPI)}.
To guarantee cross-site reproducibility and mitigate software version drift, we utilize a zero-installation execution mode that dynamically pulls PMIx-aware Apptainer containers to HPC compute nodes.
Furthermore, we demonstrate genuine model-level portability using the NESTML domain-specific language, allowing custom neuron models to be written once and automatically compiled for either the NEST (C++) or sPyNNaker backends.
Validated with a balanced random network case study, this work illustrates a practical, end-to-end path for hardware-agnostic workflows while highlighting the critical role of containerization and domain-specific languages in achieving true cross-platform reproducibility.

\todo[inline]{%
ECM: I added more high-level keywords, but I am not sure if we can put more than 6\dots{}but probably ok to fix after review.
}
\end{abstract}

\section{Introduction}
\label{sec:intro}

Modern scientific workflows increasingly span computing architectures with
fundamentally different execution models, and computational neuroscience is
a particularly clear example of this trend~\cite{amunts2024coming}. The same
spiking neural network (SNN) that a researcher wants to explore at high numerical
precision on a classical von-Neumann supercomputer may, in a follow-up
experiment, need to run in biological real-time on a digital neuromorphic
substrate such as SpiNNaker~\cite{furber2014spinnaker}, or in an accelerated
regime on a mixed-signal system such as
BrainScaleS~\cite{schmidt2023from,pehle2022brainscales2}. Each platform has
its own submission interface, its own authentication mechanism, its own
software stack with its own version constraints, and, in the worst case,
its own dialect of the network description itself.

The EBRAINS Research Infrastructure (RI) was created, in part, to flatten
this cost~\cite{amunts2024coming}. It combines a federated identity layer
based on OIDC, a curated \chadded{and versioned} software \chreplaced{environment}{stack} called the EBRAINS Software
Distribution (ESD), and a set of client libraries that expose HPC sites
(through PyUNICORE~\cite{unicore}) and the neuromorphic platform (through
the Neuromorphic Computing Platform Interface, NMPI) as programmable
\chreplaced{accessible compute resources}{services}.
In principle, this should let a researcher write a single \chadded{Jupyter}
notebook (or Python script) that dispatches the same model to a
supercomputer and to a neuromorphic \chreplaced{system}{chip} with no more than a \chreplaced{target-backend}{target-name} \chadded{parameter}
change, and do so under the FAIR principles for scientific data and
software~\cite{wilkinson2016fair}. In practice, two obstacles get in the
way. First, \chreplaced{ESD deployments are updated independently across sites}{each site re-deploys the ESD on its own schedule}, so the NEST~\cite{gewaltig2007nest}, PyNN~\cite{davison2009pynn}, or sPyNNaker~\cite{rhodes2018spynnaker} versions a user finds on a target HPC site often \chadded{differ, and might even} lag behind those in the EBRAINS JupyterLab environment---a \emph{version drift} that undermines reproducibility in ways that are easy to miss. Second,
simulator-independent APIs such as PyNN provide portability only for the
standard library of neuron models; the moment a user introduces a custom
neuron or plasticity rule, model portability breaks down and the model
must be re-implemented for each backend.

\chadded{
Beyond execution frameworks, standardizing the network representation itself remains an active area of research.
The Neuromorphic Intermediate Representation (NIR) has recently emerged as a unified format for mapping SNN topologies across disparate platforms \cite{pedersen2024neuromorphic}.
However, NIR currently focuses strictly on the static signal-flow graph (or the computational graph for numerical descriptions of dynamics), standing in contrast to PyNN, which defines a complete ``experiment flow'' including stimulation, recording, and simulation control.
Furthermore, NIR and NESTML serve complementary, rather than competing, roles in a fully hardware-agnostic stack.
While initiatives like NIRData standardize data conversion for network topologies, executing these graphs on numerical backends still requires generating efficient simulator code for individual neuron dynamics---the exact gap NESTML fills.
In the future, a highly unified workflow could naturally combine NIR for structural representation with NESTML for compiling the underlying node dynamics, all orchestrated across federated computing sites using the containerization strategies presented here.
}

Previous efforts within EBRAINS-RI and its HBP predecessor have
tackled parts of this problem. Van Albada et
al.~\cite{vanalbada2018performance} benchmarked the
Potjans--Diesmann microcircuit on NEST and SpiNNaker carefully, but
the two runs were independent and the harness was hand-built on each
side. Br\"uderle et al.~\cite{bruederle2011comprehensive} introduced
an end-to-end PyNN workflow for FACETS/BrainScaleS in 2011,
neuromorphic-only. Senk et al.~\cite{senk2017collaborative}---the
closest precedent to ours---showed in 2017 that one HBP Collaboratory
notebook could dispatch to NEST on HPC and to SpiNNaker, but did not
address version drift, which becomes the main obstacle the moment
the workflow moves to another HPC site.

In this paper, we use EBRAINS Software Distribution~(ESD) on the cloud (via JupyterLab)\footnote{\url{https://lab.ebrains.eu}}
to build and execute workflows that answer one question:
\emph{can a user dispatch the same spiking network from one notebook across multiple federated HPC sites and a neuromorphic
substrate, with the run reproducible regardless of which ESD version
each site happens to have installed, and with custom neuron models?}

\vspace{\baselineskip}
Our contributions are:
\begin{enumerate}
  \item A reference workflow, hosted entirely in EBRAINS JupyterLab, that
        targets two HPC sites through PyUNICORE, assuming the ESD \chreplaced{is}{modules are}
        pre-installed on each site (\S\ref{sec:hpc}).
 \item A \emph{pre-built container} execution mode for HPC, in which a single
      PyUNICORE submission chains two steps on the target site: a
      login-node precommand pulls a PMIx-based, ESD-derived Apptainer
      container from the EBRAINS OCI registry via ORAS, and an
      \texttt{sbatch} step then runs the simulation inside that container
      on the compute nodes (\S\ref{sec:zeroinstall}).

  \item A demonstration of the same workflow targeting both a von-Neumann
        substrate (JUSUF) and the SpiNNaker-1 neuromorphic system, via the
        Neuromorphic Computing Platform Interface (NMPI) (\S\ref{sec:nmc}).

  \item An end-to-end path through the NESTML domain-specific language, in
        which neuron and synapse models written once in the Collaboratory
        are code-generated for the appropriate target: NEST C++ on HPC, or
        the SpiNNaker toolchain on NMC (\S\ref{sec:nestml}).
\end{enumerate}

For contributions (1)--(3) we use the balanced random network~\cite{brunel2000dynamics},
a model commonly used in the neuroscience community. By changing the number
of neurons, it scales from a single SpiNNaker board to several, in the same way it scales across HPC nodes. 
Finally, we discuss what works and what does \emph{not} yet work, including the potential of Apptainer-based containers to mitigate the version drift between sites and the cloud that we observed repeatedly while building the workflow. Although this work is done in the EBRAINS JupyterLab, it
is not tied to that frontend: any client that can obtain an OIDC token from
EBRAINS-RI can drive it (\S\ref{sec:discussion}).

\section{Background: EBRAINS as a Research Infrastructure}%
\label{sec:background}

\subsection{The EBRAINS Software Distribution (ESD)}%
\label{sec:esd}
The EBRAINS Software Distribution (ESD) is the central software component
of EBRAINS Research Infrastructure:
a spack-managed ecosystem of approximately eighty top-level packages (``EBRAINS tools'').
It acts as the reproducible execution layer of EBRAINS, providing a curated and versioned software environment that enables workflows, services, and computational experiments to run consistently across heterogeneous infrastructures.
It also facilitates service deployments such as the EBRAINS JupyterLab \chreplaced{environment used in this work}{service that we use}.
At the same time, it provides a validation framework for EBRAINS tools, as its components are continuously tested together to ensure compatibility and functional interoperability across the ecosystem.
The ESD encompass HPC-enabled simulation and data 
processing tools, AI applications and workflows, access libraries to EBRAINS services 
(such as the neuromorphic platform or the neuro-robotics platform).
\chadded{Beyond these standalone tools, it supports the definition of workflow packages, which capture complete, validated sets of interoperable tools required to execute specific computational workflows.}
The ESD's components fall, roughly, into four classes (with key SNN simulation backends summarized in 
Table~\ref{tab:backends}):

\begin{table}[bt]
\centering
\caption{Simulation backends for Spiking Neural Networks (BSS-2 included
for context; not exercised in this work)}
\label{tab:backends}
\begin{tabular}{@{}llll@{}}
\textbf{Dimension} & \textbf{HPC / NEST} & \textbf{SpiNNaker-1} & \textbf{BSS-2} \\
\toprule
Arithmetic    & float64                   & fixed-point S16.15          & analog (continuous) \\
Timestep      & \SI{0.1}{ms} (tunable)    & \SI{1}{ms} (real-time)      & none (physical model) \\
Neuron model  & exact LIF ODE             & LIF, truncated              & LIF (up to MC-AdEx) \\
Time scale    & faster than real-time     & biological real-time        & $1000\times$ accelerated \\
\chreplaced{``Imprecision''}{Noise source}  & \chreplaced{---}{Poisson drive only}        & fixed-point rounding        & device mismatch \\
Parallelism   & MPI + OpenMP              & 18 cores/chip, 864/board      & 512 neurons/chip \\
Connectivity  & no constraint             & per-core memory and         & placement/routing and \\
              &                           & routing                     & calibration constraints \\
\bottomrule
\end{tabular}
\end{table}

\begin{itemize}
  \item \textbf{HPC-oriented simulators.} NEST, NEURON, and Arbor, together
        with the Python interfaces (PyNN, NEST's Python module, NESTML's
        code-generation backends) that brain modellers actually program
        against.
  \item \textbf{Hardware-specific toolchains.} BrainScaleS \chreplaced{APIs, middleware and access}{tool and
        client} libraries, SpiNNaker support packages (sPyNNaker), and
        robotics middleware used by the Neurorobotics Platform.
  \item \textbf{Visualisation and analysis.} Tools such as the siibra brain
        atlas client, NEST Desktop, and Elephant for spike-train analysis.
  \item \textbf{Services and clients.} PyUNICORE for cross-site job
        submission, the NMPI client for neuromorphic submission, and various
        Knowledge Graph and data-access libraries.
\end{itemize}

\subsection{Heterogeneous compute resources}
\label{sec:resources}

For multi-site workflows, we use JUSUF at J\"ulich
Supercomputing Centre and Galileo100 (G-100) at CINECA, 
both of which are
classical von-Neumann HPC systems with \chadded{the} ESD pre-installed
\chdeleted{-- though at
different versions: ESD 25.04 in JupyterLab, 24.10 on JUSUF, and 23.02 on
G-100}. This mismatch is one of the practical reasons we adopt the
container-based execution mode of \S\ref{sec:zeroinstall}. G-100 is used
only for the first workflow; JUSUF carries the rest of the HPC runs.
SpiNNaker-1~\cite{furber2014spinnaker} represents the neuromorphic
(non-von-Neumann) side.
Unlike standard HPC interconnects, this digital many-core substrate features a bespoke routing fabric optimized for sparse, asynchronous communication, delivering large-scale SNN simulations with exceptional energy efficiency.


\section{Workflow Design}
\label{sec:design}

The workflow is orchestrated in the JupyterLab instance hosted by the
EBRAINS Collaboratory. As a first step, the user retrieves an EBRAINS OIDC
access token from the running Collaboratory session (via
\texttt{clb\_oauth.get\_token()}); the same token is then used to
authenticate against PyUNICORE and NMPI when submitting jobs to the HPC
sites and to the neuromorphic substrate. The notebook runs under the ESD
kernel, which already provides PyUNICORE, the NMPI client, PyNN, NEST, and
NESTML. From this notebook the user assembles a workflow by specifying
three things: the \emph{network} (a script written in PyNN or NESTML), the
\emph{target} (an HPC site or SpiNNaker), and the \emph{execution recipe}
(a short Python helper that wraps the model in either a UNICORE job or an
NMPI ticket).

\subsection{Cross-site HPC submission via PyUNICORE}
\label{sec:hpc}

UNICORE~\cite{unicore} is a federation software suite that exposes HPC
compute and data resources through a RESTful API. It is deployed across
the \chreplaced{Fenix Research Infrastructure}{Fenix/ICEI} supercomputing sites, including JSC and CINECA, and
provides a uniform job-submission, data-movement, and workflow interface
on top of each site's native resource manager.
We use UNICORE through
its Python client, PyUNICORE, which is shipped as part of ESD and lets
the submission code live in the same notebook that defines the network.
A single notebook can target multiple sites by varying the registry URL
and the site name in the submission step (Listing~\ref{lst:wf1-submit});
the job description itself (Listing~\ref{lst:wf1-jobdesc}) is independent
of the site and can be reused or version-controlled.

\begin{lstlisting}[
  caption={Workflow~1 job description. Declarative, OIDC-free, and safe to
           version-control or archive as a FAIR artefact. The same
           description shape supports either of UNICORE's submission modes
           -- a batch submission to the site's scheduler, or direct
           execution on the login node -- by varying a few fields.},
  label=lst:wf1-jobdesc,
  numbers=left,
  stepnumber=1
]
# Batch-mode job: UNICORE forwards the sbatch request to
#                 the site scheduler
job_desc_batch = {
    "Executable": "sbatch",
    "Arguments":  ["run_batch.slurm"],
    "Resources": {
        "Project": "<project>",
        "Queue":   "batch",
        "Nodes":   "1",
        "Runtime": "3600", # seconds
    },
}
# Login-node job: UNICORE runs executable on the login node
job_desc_login = {
    "Executable": "/bin/bash",
    "Arguments":  ["run_login.sh"],
    "Job type":   "ON_LOGIN_NODE",
}
\end{lstlisting}

\begin{lstlisting}[
  caption={Workflow~1 submission. The OIDC token is obtained from the
           Collaboratory session and used to authenticate against the
           UNICORE registry. Either job description from
           Listing~\ref{lst:wf1-jobdesc} can be passed unchanged.},
  label=lst:wf1-submit,
  numbers=left,
  stepnumber=1
]
import pyunicore.client as uc
from pyunicore.credentials import create_credential
from clb_nb_utils import oauth as clb_oauth

credential = create_credential(token=clb_oauth.get_token())
registry   = uc.Registry(credential, "<registry-url>")
site       = registry.site("<site-name>")

job = site.new_job(
    job_description=job_desc_batch, # or job_desc_login
    inputs=["run_batch.slurm", "brunel_alpha_nest.py"],
)
job.poll() # block until DONE / FAILED
\end{lstlisting}

\subsection{Zero-installation execution with containers}
\label{sec:zeroinstall}
The common way to run a PyNN or NEST script on an HPC site is to rely
on the site's pre-installed \chreplaced{software modules}{ESD (exposed as a module)}.
While the ESD provides a curated and versioned software environment, native deployments are site-specific and, 
in practice, version availability can vary between sites:
at the time of writing, EBRAINS
JupyterLab provides \texttt{ebrains/25.04}, JUSUF provides
\texttt{ebrains/24.10}, and Galileo100 provides \texttt{ebrains/23.02}.
A script developed against one of these versions is not guaranteed to
produce reproducible results on another HPC site or in the JupyterLab
environment, which directly undermines the cross-site reproducibility
that EBRAINS RI is designed to provide.
Container-based execution mitigates \chreplaced{site-specific deployments}{this} by 
encapsulating complete runtime environments with fixed software versions. In this work, it does so by shipping a pre-built,
PMIx-aware Apptainer image with all userspace dependencies -- NEST, its
Python bindings, MPI, and their transitive dependencies -- pinned to
fixed versions.
\chadded{%
The container environment is aligned with the ESD software environment, reproducing equivalent version constraints to ensure compatibility across execution sites.
In this way, containerization complements the ESD by enabling a consistent instantiation of its curated software ecosystem across heterogeneous infrastructures, allowing workflows to be executed under controlled and reproducible conditions, independent of the specific deployment available at each site.
}%
\chadded{%
This allows workflows to be executed reproducibly even on systems where the ESD is not pre‑deployed, provided that a compatible container runtime and access mechanism (e.g. UNICORE) are available.
}%
Previous work has shown that such \chdeleted{EBRAINS HPC}
containers are portable across EuroHPC sites with negligible
performance overhead compared to bare-metal installations on CPU
clusters~\cite{singh2026hpccontainersebrainsportable}. The remaining sources of
variability are the ABI surfaces the container cannot encapsulate: the
host kernel, the resource manager (SLURM), and the MPI launcher\chadded{, and—for GPU-accelerated workloads—hardware drivers such as CUDA}.
Of these, the MPI launcher is the only one that materially affects our
workflow. NEST relies on MPI for inter-rank communication, which
requires that the container's MPI build and the host's launcher agree
on a bootstrap protocol. We rely on PMIx for this purpose: the
container ships an MPI implementation compiled against PMIx, and the
host (JUSUF) provides ParaStationMPI with PMIx support. The invocation
\texttt{srun --mpi=pspmix apptainer exec --sharens} therefore allows
the host launcher to bootstrap NEST's MPI ranks across container
instances without requiring binary compatibility between the container
and host MPI builds.

Our PyUNICORE workflow exploits this in two stages within a single
submission. The login node, which has internet access, pulls
the pre-built container from the EBRAINS Harbor \chadded{container} registry.

\begin{lstlisting}[
  caption={Workflow~2 job description. Stage~1 (\texttt{User precommand},
           executed on the login node) pulls
           the pre-built, PMIx-aware Apptainer image from EBRAINS
           Harbor. Stage~2 (the \texttt{sbatch} \texttt{Executable})
           runs the simulation on the compute nodes.},
  label=lst:wf2-jobdesc,
  numbers=left
]
job_desc_container = {
    # Stage 2: dispatched to the compute nodes after the
    #          precommand
    "Executable": "sbatch",
    "Arguments":  ["run_container.slurm"],

    # Stage 1: runs on the login node
    "Job type":   "ON_LOGIN_NODE",
    "RunUserPrecommandOnLoginNode": "true",
    "User precommand": (
        "apptainer pull -F "
        "oras://docker-registry.ebrains.eu/hpc-containers/"
        "<image>:<tag>"
    ),
    "Resources": {"Project": "<project>"},
}
\end{lstlisting}

The same
UNICORE job then dispatches an \texttt{sbatch} step to the compute
nodes, which read the image from the shared filesystem (compute nodes
have no internet access) and run the simulation under \texttt{srun
--mpi=pspmix apptainer exec --sharens}. Post-processing returns to the
login node. From the user's perspective this is one submit call;
under the hood it is a two-stage flow whose stages are stitched
together by UNICORE.

\begin{lstlisting}[
  caption={Compute-node SLURM script (\texttt{run\_container.slurm})
           dispatched by Stage~2 of the job description in
           Listing~\ref{lst:wf2-jobdesc}. The container provides NEST
           and its dependencies; the host contributes only the
           ParaStationMPI/PMIx bootstrap.},
  label=lst:wf2-slurm,
  language=bash,
  numbers=left
]
#!/bin/bash
#SBATCH --nodes=1
#SBATCH --time=01:00:00

srun --mpi=pspmix apptainer exec --sharens <image>.sif \
    python3 brunel_alpha_nest.py
\end{lstlisting}

\subsection{Neuromorphic submission via NMPI}
\label{sec:nmc}

The neuromorphic side uses a different submission interface. NMPI, the
Neuromorphic Computing Platform Interface, exposes the SpiNNaker and
BrainScaleS systems as remote services: the user uploads a Python script and
a small JSON descriptor, and the platform schedules execution on the chosen
hardware. The notebook helper for SpiNNaker is symmetrical to the UNICORE
helper:

\begin{lstlisting}[caption={Submitting the same network to SpiNNaker-1
through NMPI.}, label=lst:nmpi]
import nmpi
client = nmpi.Client(token=ebrains_oidc_token)
job_id = client.submit_job(
    source="balanced_random.py",
    platform="SpiNNaker",
    collab_id="my-collab",
    config={"spynnaker_version": "6.0.0"},)
client.wait(job_id)
result_dir = client.download_data(job_id)
\end{lstlisting}

\noindent From the user's point of view, swapping \texttt{platform="SpiNNaker"}
for a UNICORE call against JUSUF is the only line that changes between a
neuromorphic and an HPC run. The network description itself is unchanged.

\subsection{NESTML: one model, two backends}%
\label{sec:nestml}

While PyNN~\cite{davison2009pynn} abstracts the network topology and overall experiment flow, it still requires the underlying
simulator to provide the implementation for the desired neuron model. When a user wants a non-stock
neuron -- a modified adaptive exponential model, say, or a custom plasticity
rule -- the model must be implemented for each backend separately.

NESTML~\cite{plotnikov2016nestml,linssen2025nestml} was created precisely to close this last gap. It is a domain-specific language (DSL) for neuron and synapse models, paired with a code-generation toolchain that emits high-performance simulation code for a chosen target platform. The language is agnostic of numerical methods and simulation platform; the toolchain is what knows about backends. This separation supports the FAIR principles for computational models~\cite{wilkinson2016fair}, and ESD ships the NESTML toolchain together with both its NEST and SpiNNaker backends, which lets us place code generation inside the cloud notebook.
\chdeleted{%
To enable efficient and large-scale simulation of the models, the language is combined with a code generation toolchain, that, in addition to syntactically and semantically validating the models, generates high-performance code for specific platforms. The combination of a DSL with a code generation toolchain enables accessible high-performance simulation for computational neuroscience researchers and developers.}
\chdeleted{%
Additionally, a standard model interchange format like NESTML helps in making models more findable, accessible, interoperable, and reusable (``FAIR'' principles). In the context of computational models, findable means that in a database of potentially hundreds of model variants, the appropriate model can be easily found~\cite{wilkinson2016fair}. Accessible models are those that do not require extensive toolchain dependencies to work with. Interoperable models are usable across different computation hardware and simulation platforms. Reusable models are those that can be easily extended and iterated upon. These design goals are supported by the accessible, human-readable syntax of NESTML, as well as its supporting infrastructure, such as a curated model database and detailed online documentation. Writing models in NESTML makes it easy for newcomers to the field to extend and adapt models, rather than having to write low-level code or start from scratch. The wider use of NESTML as a modeling standard would facilitate interchange and promote interoperability between these software services.
ESD ships the NESTML toolchain together with its NEST and SpiNNaker code-generation backends, which means we can place the code generation and compilation step inside the cloud notebook.
}%
\chadded{%
However, bridging the gap between generated source code and a running neuromorphic simulation requires a highly specialized software stack capable of both compiling custom dynamics and mapping the user-defined network onto the physical hardware substrate.
Targeting SpiNNaker, for instance, necessitates not only an ARM cross-compiler to generate binaries for the physical cores, but also the complete SpiNNaker developer software stack to handle linking and subsequent hardware mapping and routing.
By bundling the NESTML toolchain alongside these complex, backend-specific dependencies, the ESD allows us to execute the entire code generation, cross-compilation, and mapping pipeline transparently within the cloud notebook.
}

\begin{lstlisting}[caption={Driving NESTML from the notebook to generate code
for either the NEST or SpiNNaker target.}, label=lst:nestml]
from pynestml.frontend.pynestml_frontend import generate_target

generate_target(input_path="iaf_adapt.nestml",
                target_platform="NEST",
                target_path="build/nest")
generate_target(input_path="iaf_adapt.nestml",
                target_platform="SpiNNaker",
                target_path="build/spinn")
\end{lstlisting}

The generated artefacts -- a NEST C++ module in one case, a PyNN-compatible
Python module backed by the SpiNNaker toolchain in the other -- are then
shipped with the corresponding UNICORE or NMPI job. The model itself, the
\texttt{.nestml} file, is never edited between targets. This is the closest
thing in the workflow to a genuine ``write once, run on heterogeneous
neural hardware'' experience, and in our view it is the contribution that
matters most in the long run.

\section{Case Study: A Balanced Random Network}
\label{sec:case}

To validate contributions (1)--(3) we use the balanced random network from
the sPyNNaker example collection~\cite{rhodes2018spynnaker}. The network is
a small Brunel-style configuration~\cite{brunel2000dynamics}: an excitatory
population $E$ and an inhibitory population $I$ with a 4:1 size ratio, sparse
random connectivity with probability $0.1$, Gaussian-distributed weights and
delays, and external Poisson stimulation that drives the network into an
asynchronous irregular firing regime at a few tens of Hz. The example
predates our work; we treat it as a fixed reference network and only adapt
the harness around it.

\subsection{Execution on JUSUF and Galileo100}

The PyNN script is submitted unmodified to both JUSUF and Galileo100
through the helper of Listing~\ref{lst:wf1-submit}. The simulator backend is
NEST 3.x, invoked from PyNN through \texttt{pyNN.nest}. The container image
referenced by \texttt{\$EBRAINS\_ESD\_IMAGE} was pre-built. The resulting spike trains, after the usual transient,
show the expected asynchronous irregular regime; the per-neuron firing rates
match between sites to within Monte Carlo noise of the random seed, as one
would hope.

\subsection{Execution on SpiNNaker-1}

For the SpiNNaker run, only the import line at the top of the PyNN script
changes (\texttt{import pyNN.spiNNaker as sim} instead of
\texttt{import pyNN.nest as sim}). The script is submitted via NMPI
(Listing~\ref{lst:nmpi}). sPyNNaker is responsible for mapping the network
onto SpiNNaker chips; for a Brunel network of the size shipped in the
example, this fits comfortably on a single SpiNN-5 board and executes in
real time. The recorded spikes are returned to the notebook through the NMPI
data-download endpoint and analysed with the same code path that processes
the HPC results.

We do not claim a quantitative match between the SpiNNaker run and the
NEST run; the two backends use different random-number generators, different
internal timestep semantics, and (deliberately) different floating-point
precision. What we do claim is that the \emph{workflow} is uniform: the same
notebook drove the same network, with one-line edits, against three target
machines spanning two computing paradigms, and brought all results back to
the same analysis cells.

\section{Discussion and Lessons Learned}
\label{sec:discussion}

Building the workflow we have just described turned out to be less an
exercise in writing code and more an exercise in pinning versions. We close
with three observations that we believe are useful to anyone considering a
similar effort.

\paragraph{Version drift is the dominant practical problem.}
The ESD has the explicit goal of providing identical packages everywhere, but in
practice the deployments at \chreplaced{federated}{different} sites are refreshed on different
schedules. During the work reported here we encountered combinations in which
the NEST version installed natively on a site was older than the version available in EBRAINS JupyterLab. As a result, the same script produced
different numerical outputs depending on where it ran. \chdeleted{ The container-based execution path of \S\ref{sec:zeroinstall} was, in part, a deliberate response
to this: the container bypasses the host installation entirely, so
the version a user sees in JupyterLab is the version that runs on
the compute node.}
\chdeleted{inside the ESD container, with the consequence that PyNN's auto-detection of
NEST-internal symbols silently picked the wrong one when the script ran
outside the container} .
\todo[inline,color=red]{EM/ECM4KKS: It's ``ESD module'' above, not ``ESD container'', right? The container solution is only described as a solution below?
If really both things were tried (native ESD AND containerized), this isn't clear from the text before/after; in this case, it also needs a fix somewhere.
}
\chreplaced{%
To systematically address this drift, two distinct deployment strategies emerge.
The first is the portable, HPC-enabled container execution path detailed in \S\ref{sec:zeroinstall};
by isolating the execution and forbidding the script to see any host Python, we trade native flexibility for strict cross-site reproducibility.
A second solution lies in shared software deployment frameworks like the European Environment for Scientific Software Installations (EESSI).
Distributing the ESD via EESSI would allow all federated EuroHPC sites to utilize the exact same optimized software stack, eliminating version drift without container-based encapsulation.
}{%
The container-based execution path of
\S\ref{sec:zeroinstall} was, in part, a deliberate response to this: by
forbidding the script to see any host Python at all, we trade flexibility
for reproducibility.
}%
\chreplaced{%
Similar standardization efforts are ultimately needed on the neuromorphic side, where aligning the sPyNNaker, PyNN, and NESTML backend versions still requires manual iteration.
}{%
The sPyNNaker side has its own version-coupling story -- a given sPyNNaker release expects a specific PyNN release -- and aligning both targets on a PyNN version that the NESTML SpiNNaker backend also accepts required several iterations.
}

\paragraph{Authentication is, surprisingly, the easy part.}
A single EBRAINS OIDC token grants access to every federated UNICORE site
\emph{and} to the NMPI gateway. This is a strong improvement over the
SSH-key-per-site situation that preceded EBRAINS RI, and it is the main
reason a workflow like ours can fit in a single notebook at all.

\paragraph{Genuine portability is at the model layer, not the script layer.}
PyNN gives us source compatibility for stock neuron models, which is enough
for the demonstration network. The moment a user wants a non-stock neuron,
PyNN portability evaporates and one is back to writing the model twice. The
NESTML path of \S\ref{sec:nestml} is, in our experience, the only escape:
expressing the model in a DSL whose compiler is the thing that knows about
backends. We expect that future EBRAINS work will lean further in this
direction.



\section{Conclusion}
\label{sec:conclusion}

We have presented a JupyterLab-hosted workflow \chreplaced{within}{on top of} the EBRAINS
Research Infrastructure that lets a user target multiple HPC sites and a
neuromorphic platform with the same network description and the same OIDC
identity.
\chadded{%
By leveraging the EBRAINS Software Distribution via pre-built containers and orchestrating jobs through PyUNICORE, HPC execution requires zero installation on the user's side.
}%
\chdeleted{%
Code execution on HPC requires no installation on the user's side:
the EBRAINS Software Distribution is shipped as a container, and PyUNICORE
arranges for the right invocation on each compute node.
}%
The neuromorphic
path uses NMPI symmetrically.
\chreplaced{Furthermore, to achieve true model-level portability, c}{C}ustom neuron models are
\chreplaced{defined}{written} once in
NESTML and compiled \chreplaced{directly within}{, in} the notebook\chreplaced{ for the chosen}{, for whichever} backend\chdeleted{ the user is
about to target}.
\chadded{
We successfully validated this end-to-end framework across all execution paths using a standard balanced random network.
}%
\chdeleted{%
A balanced random network from the sPyNNaker example
collection is enough to exercise all three paths.
}%

\chreplaced{%
Despite these advances, practical deployment challenges remain.
}{%
We are aware of the limits of the current setup.
}%
\chadded{Software} version drift between \chadded{site-specific} ESD deployments is the dominant source of friction.
\chreplaced{%
Future work must address this gap,
}{%
, and we would like to see future work close that gap --
}%
either by tightening the release-engineering loop or by
\chreplaced{%
}{%
providing
}{%
making
}%
the container-based path the default \chreplaced{across all}{on every} EBRAINS site\chadded{s}.
\chadded{%
A longer-term solution would be the integration of the ESD software environment into shared frameworks like EESSI.
}%
We would also like to push more of the model layer
into NESTML, including plasticity rules, so that the ``write once'' property
of \S\ref{sec:nestml} extends to a wider class of experiments.

\begin{credits}
\subsubsection{\ackname}%
This work has received funding from
the EC Horizon 2020 Framework Programme
under grant agreement 
945539 (HBP SGA3), 
the EC Horizon Europe Framework Programme
under grant agreement
101147319 (EBRAINS 2.0).
\subsubsection{\discintname}%
The authors have no competing interests to declare that are relevant to the content of this article.
\end{credits}

\bibliographystyle{splncs04}
\bibliography{references}
\end{document}